\documentclass[sigconf,nonacm]{acmart}
\usepackage{enumitem}
\usepackage{multirow}

\AtBeginDocument{%
	\providecommand\BibTeX{{%
			\normalfont B\kern-0.5em{\scshape i\kern-0.25em b}\kern-0.8em\TeX}}}

\copyrightyear{2022} 
\acmYear{2022} 
\setcopyright{acmcopyright}\acmConference[CIKM '22]{Proceedings of the 31st International ACM International Conference on Information and Knowledge Management}{October 17--22, 2022}{Atlanta, Georgia, USA}
\acmBooktitle{Proceedings of the 31st International ACM International Conference on Information and Knowledge Management (CIKM '22), October 17--22, 2022, Atlanta, Georgia, USA}
\acmPrice{15.00}
\acmDOI{xxx}
\acmISBN{xxx}

\begin{document}
	\fancyhead{}
	\title{Soft Retargeting Network for Click Through Rate Prediction}
	
	
	
	 \author{Xiaochen Li}
	 \affiliation{%
		   \institution{Alibaba Group}
		   \city{Beijing}
		   \country{China}
		 }
	 \email{xingke.lxc@alibaba-inc.com}
	
	 \author{Xin Song}
	 \affiliation{%
		   \institution{Alibaba Group}
		   \city{Beijing}
		   \country{China}
		 }
	 \email{songxin.song@alibaba-inc.com}
	
	 \author{Pengjia Yuan}
	 \affiliation{%
		   \institution{Alibaba Group}
		   \city{Beijing}
		   \country{China}
		 }
	 \email{pengjia.ypj@alibaba-inc.com}
	
	 \author{Xialong Liu}
	 \affiliation{%
		   \institution{Alibaba Group}
		   \city{Beijing}
		   \country{China}
		 }
	 \email{xialong.lxl@alibaba-inc.com}
	
	 \author{Yu Zhang}
	 \affiliation{%
		   \institution{Lazada Group}
		   \city{Beijing}
		   \country{China}
		 }
	 \email{daoji@lazada.com}
	
	\begin{abstract}
		 The study of user interest models has received a great deal of attention in click through rate (CTR) prediction recently. These models aim at capturing user interest from different perspectives, including user interest evolution, session interest, multiple interests, etc. In this paper, we focus on a new type of user interest, i.e., user retargeting interest.  User retargeting interest is defined as user's click interest on target items the same as or similar to historical click items.  We propose a novel soft retargeting network (SRN) to model this specific interest. Specifically, we first calculate the similarity between target item and each historical item with the help of graph embedding. Then we learn to aggregate the similarity weights to measure the extent of user's click interest on target item. Furthermore, we model the evolution of user retargeting interest. Experimental results on public datasets and industrial dataset demonstrate that our model achieves significant improvements over state-of-the-art models. 
	\end{abstract}

	\ccsdesc[500]{Information systems~Online advertising}
	
	\keywords{Click-Through Rate Prediction; User Behavior Modeling; User Retargeting Interest}
	
	\maketitle
	
	\section{Introduction}
	Click through rate prediction plays a vital role in recommender system and online advertising. Due to the rapid growth of user historical behavior data, user behavior modeling has been widely adopted in CTR prediction, which focuses on capturing the dynamics of user interest from user historical behaviors~\cite{DIN,DIEN,2020ComicRec,bert4rec,DSIN,DeepForCtr,BST,MIMN,SIM,HPMN}. These works model user interest from different perspectives, including capturing user interest using attention~\cite{DIN} or transformer ~\cite{BST,bert4rec} , user interest evolution~\cite{DIEN}, user interest interaction among sessions~\cite{DSIN}, multiple interests~\cite{2020ComicRec}, and long-term user behavior~\cite{MIMN,SIM,HPMN}.

	In this paper, we aim to model a specific user interest, i.e., \textit{user retargeting interest}. User retargeting interest is user's click interest on the items the same as or similar to historical clicked items (we name them \textit{retargeted items}). These items are usually retrieved by retargeting service~\cite{Kantola2014} for reminding users of the products they visited before, or by matching stage of online advertising system for recommending similar products based on user historical behavior. As retargeted items are highly relevant to user's recent interest, they can improve user engagement with products/brands dramatically in online advertising. It is reported that the average CTR of retargeted ads is 10 times higher than regular display ads~\cite{ret_higher}. We also observe a similar phenomenon in our native ads scenario that the CTR of retargeted items is 2 times higher than other items. Although retargeted items are much more likely to be clicked, conventional user behavior models do not distinguish retargeted items from other items, which may underestimate the CTR of retargeted items and lose the opportunity of getting more clicks and revenue.

	
	To understand user retargeting interest, we examine the cases of target items and user behavior sequences in our scenario. We notice that retargeted item has a strong correlation to user historical click items. For a retargeted item, it is obvious that the user has clicked the same or similar items before and left corresponding click records. Our idea is to locate and make good use of these records for representing user retargeting interest.
	
	We propose a novel soft retargeting network (SRN) to model user retargeting interest.  We first calculate similarity weights between target item and historical items with the help of graph embedding. Then we aggregate the similarity weights to measure the extent of user's click interest on retargeted item. Furthermore, we model the evolution of user retargeting interest. To the best of our knowledge, our study is the first to model user retargeting interest in click through rate prediction. 

	The rest of the paper is organized as follows: Section 2 introduces the SRN model. Section 3 presents experimental results on public datasets and industrial dataset. Section 4 concludes the paper.
	
	\section{The proposed approach}
	In this section, we first introduce a hard retargeting network (HRN) as a naive approach to model user retargeting interest. We then elaborate the network structure and key components of SRN.
	
	\subsection{Hard retargeting network}
	To measure user's interest on retargeted item, the basic idea of HRN is to count how many times user has clicked this item in user behavior sequence and use the click count to represent user interest. Intuitively, the more user clicks an item before, the more likely user is interested in the item.
	
	HRN first enumerates the items in user behavior sequence and checks whether target item and historical item are the same item by comparing their embedding vectors:
	\begin{equation}
		\label{eqn:similarity_weight}
		s(t,b_{j}) = \left\{
			\begin{array}{ll}
				1, & if\ e_{b_{j}}=e_{t}, \\
				0, & else.
			\end{array}
		\right.
	\end{equation}
	where $e_{t}$ is the embedding of target item $t$, $e_{b_{j}}$ is the embedding of historical item $b_{j}$, $s(i,j) \in [0,1]$ is the \textit{similarity weight} between target item $i$ and historical item $j$. Note that target item and historical item share the same embedding dictionary, we can ensure two items are the same if their embedding vectors are equal.
	
	For user behavior sequence $B=\{b_{1}, b_{2}, …,b_{n}\}$, we can get a sequence of similarity weights $S_{t}=\{s(t,b_{1}), s(t,b_{2}), …,s(t,b_{n})\}$. $S_{t}$ can be viewed as user's historical click records on target item. 
	
	We perform max pooling on $S_{t}$ and get $s_{t}^{max}=max_{j}(s(t,b_{j}))$ for target item $t$. If $s_{t}^{max}=1$, target item is a retargeted item for HRN. In addition, we perform sum pooling on $S_{t}$ to get:
	\begin{equation}
		\label{eqn:click_number}
		N_{S} = \sum_{j=1}^{n}s(t,b_{j})
	\end{equation}
	where $N_{S} \in [0,n]$ corresponds to the historical click number of target item. We then adopt binning method to transform $N_{S}$ into a categorical feature $FEA_{I_{S}}=binning(N_{S}, 1)$, where $bining(x, z)$ is an equal width interval binning method~\cite{binning} to discrete $x$ with bin size $z$. For example, if $N_{S}=5.0$, $FEA_{I_{S}}=``6"$. HRN looks up the embedding of $FEA_{I_{S}}$ from embedding layer and feeds it into CTR model for prediction. Although simple, the performance of HRN is particularly good in e-commerce scenarios. 
	 
	 For simplicity, we only consider historical click item sequence during the introduction of HRN and SRN. Our approach can easily be extended to historical click brand/shop/category sequence. 
	
	\subsection{Soft retargeting network}
	In HRN, only target item the same as historical item is treated as retargeted item. The ratio of retargeted items in sample data is rather low, e.g., only 5\% for item and 15\% for shop in our scenario, which heavily limits the performance of HRN. To overcome the limitation, we propose a SRN model to expand the scope of retargeted items and model user retargeting interest at a broader level.
	
	The framework of SRN is illustrated in Fig.~\ref{fig:srn_network}. The idea of SRN is similar to that of HRN. It computes a sequence of similarity weights for user behavior sequence and aggregates similarity weights to represent user retargeting interest. We will detail the key components of SRN in the following subsections:
	
	\begin{figure}[h]
		\centering
		\includegraphics[width=0.9\linewidth]{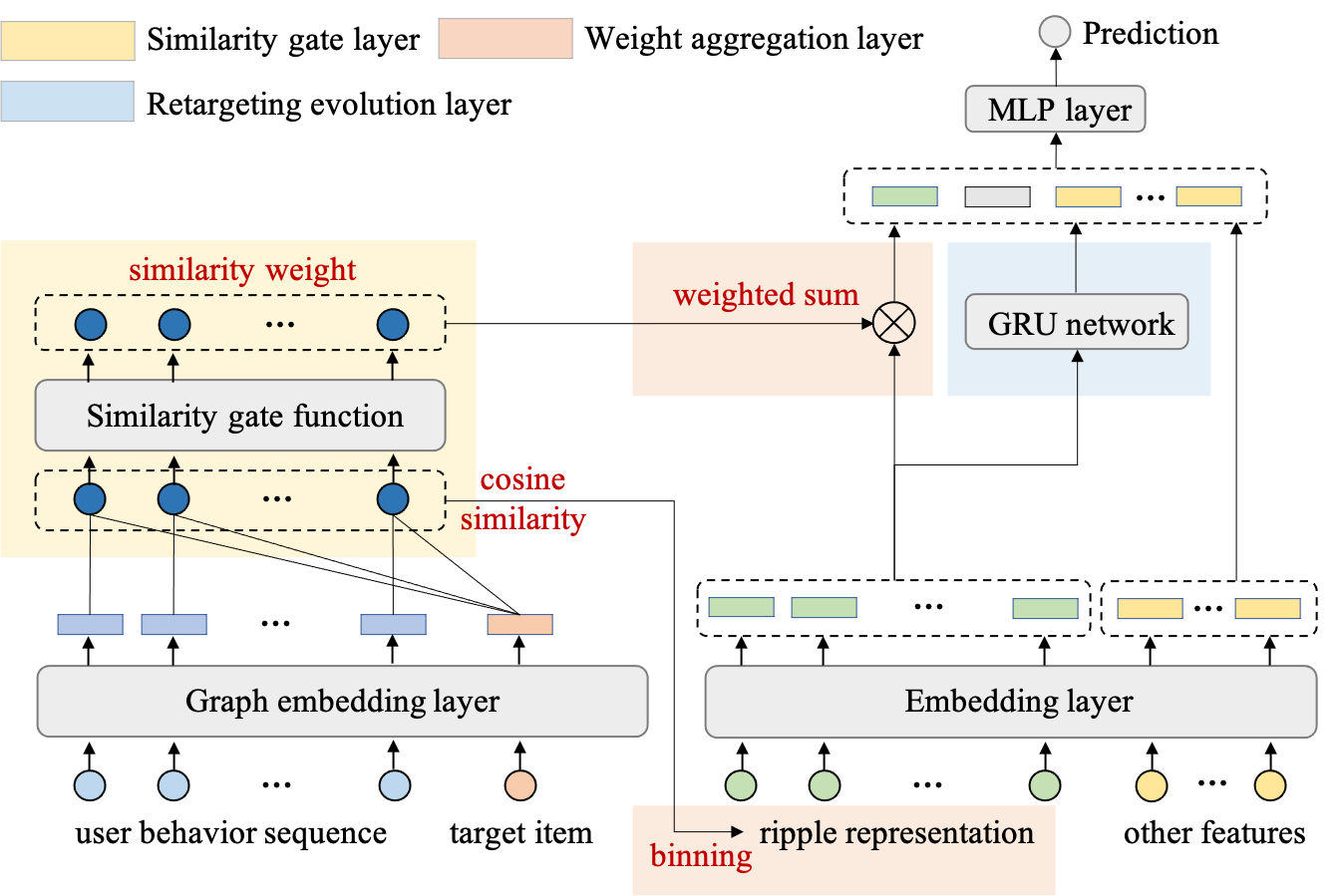}
		\caption{The framework of SRN model. Graph embedding layer represents target item and historical item with graph embedding. Similarity gate layer calculates cosine similarities between target item and historical items and employs similarity gate function to get similarity weights. Weight aggregation layer discretizes similarity weights to get ripple representation and represents user retargeting interest by aggregating similarity weights. Retargeting evolution layer uses GRU to model the evolution of user retargeting interest.}
		\label{fig:srn_network}
	\end{figure}

	{\bfseries Graph embedding layer.} To measure the similarity between target item and history items, a natural choice is just using the item embedding learned by CTR model to calculate cosine similarity. However, we find that the quality of CTR embedding is poor in practice. It is probably due to the fact that CTR embedding is learned for CTR tasks and may be incapable of capturing the similarity between items. Inspired by the success of graph embedding~\cite{graph_survey,HGAN,inductive_graph,GCNN}, we switch to using graph embedding to represent items in SRN. We pre-train a graph network based on user-item interaction records and generate a graph embedding dictionary $GE = [ge_{1}, ge_{2}, …, ge_{M}] \in R^{D \times M}$ (see Section 3.1 for more details). Graph embedding layer looks up $GE$ and generates a graph embedding $ge_{t}$ for target item $t$ and a graph embedding $ge_{b_{j}}$ for each historical item $b_{t_{j}}$.
	
	{\bfseries Similarity gate layer.} Based on graph embedding, similarity gate layer calculates the cosine similarity between target item $t$ and historical item $b_{j}$:
	\begin{equation}
		\label{eqn:cos_sim}
		cosine(t,b_{j}) = \frac{ge_{t}^{T}ge_{b_{j}}}{\lVert ge_{t} \rVert \cdot \lVert ge_{b_{j}} \rVert}
	\end{equation}
	 
	 Note that cosine similarity is different from the similarity weight used in HRN. Firstly, $cosine(t,b_{j})\in[-1,1]$ but $s(t,b_{j})\in[0,1]$. Secondly, cosine similarity is a metric used to measure the similarity between embedding vectors. Similarity weight is item similarity visually perceived by user. While there is a strong correlation between them, there are also many cases where they differ.
	 
	 We carefully examine items in our scenario and summarize three correlation rules for cosine similarity and similarity weight:
	 \begin{itemize}[leftmargin=*]
	 	\item {\bfseries Rule 1}: When $cosine(t,b_{j})=1.0$, item $t$ and item $b_{j}$ are the same item and $s(t,b_{j})=1.0$.
		\item  {\bfseries Rule 2}: When $cosine(t,b_{j})\le T$ (e.g., $T=0.90$), item $t$ and item $b_{j}$ can be two different products, e.g., iphone and microphone. In this case, $s(t,b_{j})$ drops dramatically and should approach $0.0$.
		\item  {\bfseries Rule 3}: When $T< cosine(t,b_{j})<1.0$, item $t$ and item $b_{j}$ can be the same product sold by different shops, or two products with the same brand, e.g., iPhone 11 and iPhone SE. In this case, $s(t,b_{j})$ should be a large value, e.g., 0.95.
	 \end{itemize}

 	 To satisfy these rules, we design a similarity gate function $F(x)$ to map $cosine(t,b_{j})$ to $s(t,b_{j})$:
 	 
	\begin{figure}[h]
	\centering
	\includegraphics[width=0.9\linewidth]{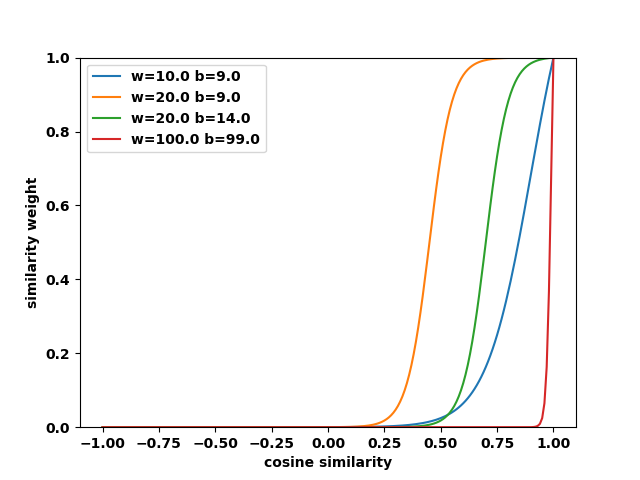}
	\caption{The curve of similarity gate function.}
	\label{fig:similarity_gate_func}
	\end{figure}

	\begin{equation}
		\label{eqn:scaled_sigmoid}
		F(x) = \frac{\sigma(w\cdot x-b)}{\sigma(w-b)}
	\end{equation}

	\begin{equation}
		\label{eqn:soft_similarity_weight}
		s(t,b_{j}) = F(cosine(t,b_{j}))=\frac{\sigma(w\cdot cosine(t,b_{j})-b)}{\sigma(w-b)}
	\end{equation}
	where $w>0$ and $b>0$. Fig. ~\ref{fig:similarity_gate_func} plots the curve of $F(x)$ for different values of $w$ and $b$. We can see $F(x)$ indeed suppresses $s(t,b_{j})$ when $cosine(t,b_{j})$ is below some threshold $T$ and the value of $T$ is determined by $b$. When $cosine(t,b_{j}) \in (T, 1)$, $F(x)$ outputs a high value of $s(t,b_{j})$ and the shape of $F(x)$ in  $(T, 1)$ is controlled by $w$. In practice, we usually set good initial values for $w$ and $b$ (e.g., $w=10.0$ and $b=9.0$) and let CTR model learn the optimal $F(x)$.   

	For user behavior sequence, similarity gate layer outputs a sequence of similarity weights $S_{t}=\{s(t,b_{1}), s(t,b_{2}), …,s(t,b_{n})\}$.

	{\bfseries Weight aggregation layer.}  We perform max pooling on $S_{t}$ to get the maximum similarity weight $s_{t}^{max}$ just like in HRN. $s_{t}^{max}$ reflects the likelihood that target item $t$ is a retargeted item. The higher the value of $s_{t}^{max}$, the more likely $t$ is a retargeted item. We can see that the scope of retargeted item in SRN is definitely broader than that in HRN.
	
	We can follow the operation in HRN to model user retargeting interest, i.e., sum pooling on $S_{t}$ and binning. However, the binning operation is non-differentiable and blocks the updating of similarity gate layer. Instead, we put binning operation on $cosine(t,b_{j})$ and generate a categorical feature $FEA_{j}^{ripple}=binning(cosine(t,b_{j}), 0.01)$ (we name it \textit{ripple representation}). We look up the embedding vector of ripple representation from embedding layer as $e_{j}^{ripple}$ and represent user retargeting interest as:
	\begin{equation}
		\label{eqn:weight_aggregation}
		I_{S} = \sum_{j=1}^{n}s(t,b_{j}) \cdot e_{j}^{ripple}
	\end{equation}

	As $F(x)$ is a linear function with limited model capacity, $s(t,b_{j})$ may be inaccurate when $cosine(t,b_{j}) \in (T, 1)$. Ripple representation can complement the limitation of $F(x)$ by providing an embedding representation of $cosine(t,b_{j})$.
	
	{\bfseries Retargeting evolution layer.} We notice that user retargeting interest may evolve over time. Suppose there are two sequences of similarity weights $S_{1}=\{0.1, 0.5, 0.9\}$ and $S_{2}=\{0.9, 0.5, 0.1\}$, weight aggregation layer would produce the same results. However, $S_{1}$ suggests that user's interest on target item grows rapidly whereas $S_{2}$ indicates that user may lose interest in target item. We use GRU network~\cite{GRU_network} to capture the evolution of user retargeting interest:
	\begin{equation}
		\label{eqn:retargeting_gru}
		G_{S} = GRU({e_{1}^{ripple}, e_{2}^{ripple}, ..., e_{n}^{ripple}})
	\end{equation}

	We concatenate $I_{S}$ and $G_{S}$ to get the output of SRN $I_{SRN}=[I_{S};G_{S}]$ and feed $I_{SRN}$ into MLP layer for CTR prediction.
	
	\section{Experiments}
	In this section we present the experimental setup and conduct experiments to evaluate the performance of our model.
	
	\subsection{Experimental setup}
	
	{\bfseries Datasets.}
	As SRN is mainly designed for click interest, we choose two public datasets which contain user click records and an industrial dataset for evaluation: 1) {\bfseries Taobao dataset\footnote{\url{https://tianchi.aliyun.com/dataset/dataDetail?dataId=649}}}. It is a collection of user behaviors from Taobao’s recommender system~\cite{zhu2018learning} and contains 89 million records, 1 million users, and 4 million items from 9407 categories. We only use click behaviors of each user. 2) {\bfseries Alimama dataset\footnote{\url{https://tianchi.aliyun.com/dataset/dataDetail?dataId=56}}}. It is a dataset of click rate prediction about display ads provided by Alimama, an online advertising platform in China.  It contains 26 million display/click records from 20170506 to 20170513, 1 million users, and 0.85 million items. We use the first 7 days' data as training sample and the last day’s data as test sample. 3) {\bfseries Industrial dataset}. It is collected from our online advertising system for a native ads scenario from 20220318 to 20220417. The dataset contains 780 million display/click records, 10.4 million users, and 0.78 million items. Each record contains a real-time click sequence from the preceding 3 days. We use the first 30 days' data as training sample and the last day's data as test sample.
	
	{\bfseries Graph Construction.} For each dataset, we pre-train a heterogeneous graph network based on the training sample of CTR task. The types of graph nodes include user, item, and its side information (brand/shop/category for Alimama dataset and industrial dataset, and category for Taobao dataset). The graph edges include: 1) {{\bfseries user-item edge}}. If user $u$ clicks item $i$, there is an edge between $u$ and $i$. User-item edge is used as label in link prediction task. 2) {{\bfseries user-side information edge}}. If user $u$ clicks an item with side information $v$ (e.g., shop), there is an edge between $u$ and $v$. 3) {{\bfseries item-item edge}}. If item $i$ and item $j$ are adjacent in user behavior sequence and the time interval between item $i$ and item $j$ is within 60 seconds,  there is an edge between $i$ and $j$. 4) {{\bfseries item-side information edge}}. If item $i$ has a side info $v$, there is an edge between $i$ and $v$.
	
	Notice that the edges are undirected and unweighted. We then traverse the graph to sample nodes and their neighborhoods according to meta paths. Meta paths include $user \rightarrow brand/shop/category \rightarrow item$, $item \rightarrow item$, $item \rightarrow brand/shop/category \rightarrow item$,  $brand \rightarrow item$,  $shop \rightarrow item$,  $category \rightarrow item$. We aggregate the node's neighborhoods using the HAN approach~\cite{HGAN} and adopt a link prediction task to supervise the learning of embedding.
	
	{\bfseries Competitors.} We compare SRN with HRN and some widely used CTR models, including DNN, DIN~\cite{DIN}, DIEN~\cite{DIEN} and BST~\cite{BST}.
	
	{\bfseries Parameter Configuration.}
	Alimama and industrial dataset use historical click item/brand/shop/category sequence. Taobao dataset uses historical click item/category sequence. The maximum sequence length is 100 and 64 for public datasets and industrial dataset respectively. The dimension of CTR embedding and graph embedding are 8 and 32 respectively. The optimizer is Adagrad with learning rate 0.01. For SRN, $w$, $b$ is 10, 9 for historical item sequence and 10, 8 for historical brand/shop/category sequence, respectively.
	
	{\bfseries Evaluation Metrics.} We use two common used metrics, $AUC$ and $LogLoss$, to evaluate these models.
	
	\begin{table}
		\caption{Results on public datasets and industrial dataset}
		\label{tab:public and industrial result}
		\begin{tabular}{l|c|c|c|c|c|c}
			\hline
			\multirow{2}{*}{Models} & \multicolumn{2}{|c|}{Taobao} & \multicolumn{2}{|c|}{Alimama} &
			\multicolumn{2}{|c}{Industrial} \\
			\cline{2-7}
			& AUC & LogLoss & AUC & LogLoss & AUC & LogLoss \\
			\hline
			DNN & 0.8557 & 0.3254 & 0.6359 & 0.1941 & 0.6805 & 0.1244 \\
			DIN & 0.8731 & 0.3080 & 0.6378 & 0.1938 & 0.6838 & 0.1243 \\
			DIEN & 0.8623 & 0.3200 & 0.6369 & 0.1939 & 0.6864 & 0.1240 \\
			BST & 0.8573 & 0.3240 & 0.6401 & 0.1936 & 0.6876 & 0.1238 \\
			HRN & 0.8838 & 0.2919 & 0.6376 & 0.1938 & 0.6842 & 0.1241 \\
		    SRN & \textbf{0.8853}& \textbf{0.2905}& \textbf{0.6609}& \textbf{0.1904}& \textbf{0.6918}& \textbf{0.1235}\\ 
		    \hline
		\end{tabular}
	\end{table}

	\begin{table}
		\caption{Retargeting ratio for HRN \& SRN}
		\label{tab:retargeting_ratio}
		\begin{tabular}{c|c|c|c|c}
			\hline
			Models & Item & Brand & Shop & Category \\
			\hline
			HRN & 5.43\% & 11.3\% & 14.9\% & 47.5\%  \\
			SRN ($s_{t}^{max}>0.50$) & 33.4\% & 13.3\% & 26.3\% & 51.0\%  \\
			\hline
		\end{tabular}
	\end{table}

	\begin{table}
		\caption{AUC for retargeted items and the others}
		\label{tab:AUC_retargeted}
		\begin{tabular}{c|c|c}
			\hline
			Models & Retargeted items & The others \\
			\hline
			DNN & 0.6748 & 0.6791 \\
			DIEN & 0.6816 (+0.0058) & 0.6861 (+0.0070)  \\
			BST & 0.6809 (+0.0061) & 0.6847 (+0.0056) \\
			SRN & 0.6914 (+0.0166) & 0.6881 (+0.0090) \\
			\hline
		\end{tabular}
	\end{table}

	\begin{table}
		\caption{Average similarity for intra- and inter-category}
		\label{tab:emb_quality}
		\begin{tabular}{c|c|c}
			\hline
			Average similarity & CTR embedding & Graph embedding  \\
			\hline
			Intra-category & 0.03294 & 0.5413  \\
			Inter-category & 0.03285 & 0.0094  \\
			\hline
		\end{tabular}
	\end{table}

	\begin{table}
		\caption{Ablation study}
		\label{tab:ablation study}
		\begin{tabular}{l|l|l}
			\hline
			Model & AUC & Diff\\
			\hline
			$SRN$ & 0.6918 & \\
			$SRN\ w/o\ GRU $ & 0.6909 & -0.0009 \\
			$SRN\ w/o\ GE $ & 0.6876 & -0.0042 \\
			$SRN\ w/o\ SIM\_GATE $ & 0.6905 & -0.0013 \\
			$SRN_{binning}$ & 0.6886 & -0.0032 \\
			\hline
		\end{tabular}
	\end{table}
	
	\subsection{Performance evaluation}	
	As shown in Table~\ref{tab:public and industrial result}, HRN achieves solid performance gain compared to DNN base. SRN outperforms HRN with an AUC gain of 0.0015, 0.0233, 0.0076 in public datasets and industrial dataset, respectively. To explain the advantages of SRN over HRN, we assume target item with $s_{t}^{max}>0.5$ is retargeted item in SRN and define \textit{retargeting ratio} for item as the ratio of the total number for retargeted items to sample number in test dataset. Retargeting ratio for brand/shop/category can be defined in the same way. We compare the retargeting ratio of SRN with that of HRN in industrial dataset. Table~\ref{tab:retargeting_ratio} shows the retargeting ratio of SRN is much higher than HRN for item and shop. It seems that SRN can capture more retargeted items/shops and thus boost the performance significantly.
	
	SRN outperforms DNN, DIN, DIEN, and BST significantly with an AUC gain of at least 0.0122, 0.0208, 0.0042 in the three datasets, respectively. To demonstrate the effectiveness of SRN in modeling user retargeting interest, we compare the performance of SRN with that of baseline models on retargeted items (also selected by $s_{t}^{max}>0.5$) and the other items in industrial dataset. Table~\ref{tab:AUC_retargeted} shows that the AUC gain of SRN on retargeted items is much higher than that on the others. For DIEN and BST, the AUC gain on retargeted items is nearly the same as that on the others. 
	
	We conduct online A/B experiments to evaluate SRN in online system. The experiment lasts for 14 days and SRN achieves 6.88$\%$ CTR and 6.12$\%$ RPM gain compared to product model (DIN+HRN).
	
	\subsection{Ablation study}
	The result of ablation study on industrial dataset is shown in Table~\ref{tab:ablation study}. Retargeting evolution layer (GRU) brings a notable AUC gain of 0.0009. $SRN\ w/o\ SIM\_GATE $ sets $s(t,b_{j})$ to $1.0$ and relies on ripple representation to represent the interest $I_{S}$, which causes an AUC loss of 0.0013. It demonstrates the importance of similarity weights in capturing user retargeting interest. $SRN_{binning}$ adopts the summing and binning operation used in HRN to aggregate similarity weights, which causes an AUC loss of 0.0032. This suggests that binning operation may lead to sub-optimal performance and weight aggregation layer is more flexible and powerful.
	
	Using CTR embedding instead of graph embedding ($SRN\ w/o\ GE $) causes the performance to decrease dramatically, which indicates the quality of graph embedding is much better than CTR embedding. To evaluate the quality of graph/CTR embedding, we select top 100 categories according to PV number and sample 100 items for each category in industrial dataset. We calculate the cosine similarities between each item and the other items using graph/CTR embedding respectively. For each item, we can get the average similarity within the same category (\textit{intra-category similarity}) and from the different categories (\textit{inter-category similarity}). We then compute the average intra- and inter-category similairity for graph/CTR embedding respectively. Table~\ref{tab:emb_quality} shows that intra-category similarity is much higher than inter-category similarity for graph embedding whereas the difference is negligible for CTR embedding. It demonstrates the superiority of graph embedding in identifying similar items.

	\section{Conclusion}
	In this paper, we propose a novel soft retargeting network to model user retargeting interest. SRN calculates the similarity between target item and historical click items with graph embedding and generates user's historical click records on items similar to target item. It then aggregates these records to represent user's interest on retargeted items. The experimental results demonstrate that our model achieves significant improvements over the competitors.
	
	\bibliographystyle{ACM-Reference-Format}
	\bibliography{cikm22}
	
\end{document}